\documentclass[sigconf]{acmart}
\usepackage{adjustbox}
\usepackage{multirow}
\usepackage{caption}
\usepackage{subcaption}

\setcopyright{none}
\renewcommand\footnotetextcopyrightpermission[1]{} 
\AtBeginDocument{%
  \providecommand\BibTeX{{%
    \normalfont B\kern-0.5em{\scshape i\kern-0.25em b}\kern-0.8em\TeX}}}

\begin{document}

\title{Semi-supervised Learning of Perceptual Video Quality by Generating Consistent Pairwise Pseudo-Ranks}


\author{Shankhanil Mitra}
\affiliation{%
 \institution{Indian Institute of Science}
 \streetaddress{C.V.Raman Road}
 \city{Bangalore}
 \state{Karnataka}
 \country{India}}

\author{Saiyam Jogani}
\affiliation{%
 \institution{BITS Pilani}
 \city{K.K.Birla, Goa Campus}
 \state{Goa}
 \country{India}}
 
\author{Rajiv Soundararajan}
\affiliation{%
 \institution{Indian Institute of Science}
 \streetaddress{C.V.Raman Road}
 \city{Bangalore}
 \state{Karnataka}
 \country{India}}

\renewcommand{\shortauthors}{Mitra, et al.}

\begin{abstract}
Designing learning-based no-reference (NR) video quality assessment (VQA) algorithms for camera-captured videos is cumbersome due to the requirement of a large number of human annotations of quality. In this work, we propose a semi-supervised learning (SSL) framework exploiting many unlabelled and very limited amounts of labelled authentically distorted videos. Our main contributions are two-fold. Leveraging the benefits of consistency regularization and pseudo-labelling, our SSL model generates pairwise pseudo-ranks for the unlabelled videos using a student-teacher model on strong-weak augmented videos. We design the strong-weak augmentations to be quality invariant to use the unlabelled videos effectively in SSL. The generated pseudo-ranks are used along with the limited labels to train our SSL model. Our primary focus in SSL for NR VQA is to learn the mapping from video feature representations to the quality scores. We compare various feature extraction methods and show that our SSL framework can lead to improved performance on these features. In addition to the existing features, we present a spatial and temporal feature extraction method based on predicting spatial and temporal entropic differences. We show that these features help achieve a robust performance when trained with limited data providing a better baseline to apply SSL. 
Extensive experiments on three popular VQA datasets demonstrate that a combination of our novel SSL approach and features achieves an impressive performance in terms of correlation with human perception, even though the number of human-annotated videos may be limited.
\end{abstract}



\maketitle

\section{Introduction}
The ubiquitous availability of mobile cameras has led to a proliferation in the generation of video content. Numerous videos are captured by humans  worldwide everyday and uploaded and shared through video service providers. With the rapid growth of such content, it becomes essential to monitor and control the quality of the captured videos for efficient storage, transmission, and retrieval. This motivates the study of perceptual video quality assessment (VQA) algorithms to provide video quality scores according to human judgements. 

VQA algorithms can be broadly classified into full reference (FR), reduced reference (RR) and no reference (NR) methods. FR methods require a reference video for comparison to evaluate the quality of a distorted video \cite{movie, stmad, ms_ssim}. On the other hand, RR methods require a small amount of information from the reference for quality assessment of the distorted video \cite{strred}. NR methods operate only on the distorted video and do not require any reference video for comparison \cite{vbliind}. In this work, our primary focus is on the VQA of camera captured videos that are authentically distorted during the capture process and where a reference video is usually not available. Thus, the NR VQA setting is most relevant in such scenarios.  

NR VQA algorithms are popularly designed in a machine learning framework. The algorithms are developed by extracting video features and regressing them against human opinion scores obtained through subjective studies. Indeed, deep networks are increasingly being studied \cite{vmeon, deep3dcnnvqa, qa_in_the_wild} for the design of NR VQA algorithms. Nevertheless, such approaches require a large amount of human annotated videos with quality scores. Since the conduct of large scale human studies to collect such scores can be pretty cumbersome, such an approach does not scale when newer studies need to be conducted as more and more diversely distorted videos are generated. The human annotation of video quality is also much more time consuming when compared with image quality or other annotation tasks such as image classification. This motivates the study of NR VQA algorithms with few labelled videos. Thus we focus on the problem of designing learning based NR VQA algorithms with limited labels. In addition to the limited labels, we also assume access to unlabelled videos to design our VQA models. Thus, one could view our problem as a \emph{semi-supervised} NR VQA problem. 

The problem of semi-supervised NR image or video quality assessment has been hardly studied in literature to the best of our knowledge. Although there exists some work on semi-supervised image quality assessment \cite{semi_iqa, semi_iqa_ensemble, sslIQA}, their extension to video is non-trivial.  Further, no studies investigate the training of video quality models with limited labelled data. While several strategies for semi-supervised learning have been explored in image and video classification, many of those strategies are not directly applicable to NR VQA. For example, consistency regularization methods \cite{consistancy1} are usually based on quality degradation of the image/video and are thus not appropriate for VQA. Many of the pseudo-labelling strategies are designed for classification \cite{pseudolabel}, and their direct application to VQA is also not obvious. 

We present a novel and reliable semi-supervised learning (SSL) strategy for NR VQA. We perform semi-supervised learning by enforcing consistency regularization on the unlabelled examples. A popular approach to achieve consistency regularization is through student-teacher networks where the student and teacher models are required to make consistent predictions on augmented unlabelled data. However, the augmentations studied in literature generally tend to distort the video quality thereby rendering such approaches irrelevant for VQA. One of our main contributions is in the design of quality invariant strong-weak augmentations that enable us to apply consistency regularization for semi-supervised NR VQA. The teacher model predictions are usually considered as pseudo-labels for the unlabelled data using which the student model can be updated.  In earlier works on consistency regularization \cite{ladder_network}, the unreliability of the pseudo-labels limits the learning of the student network, which is also termed as confirmation bias. To account for the unreliability of the pseudo-labels, we hypothesize that when the quality predictions of a pair of unlabelled videos differ beyond a threshold, their pairwise ranking is likely correct. Thus, we train the student and teacher networks to be consistent in their  pairwise quality rank predictions of video pairs.



 Although we can apply our SSL strategy to different feature extraction methods to yield improvements, we observe that learning good video quality features upfront before performing SSL on the target dataset can yield more reliable pseudo-labels. In recent years, successful CNN based methods on synthetically distorted videos were designed \cite{heke-csvt,nrsted} by learning on weak quality labels such as GMSD \cite{gmsd}, MS-SSIM \cite{ms_ssim}, ST-RRED \cite{strred}.
 Thus, we design a video quality feature learning method on videos suffering from synthetic distortions such as compression and transmission losses. In particular, we learn deep spatial, and temporal video quality features using the spatio-temporal entropic differences on video frames. Thus our approach does not use human labels while learning quality features during this pre-training. The feature extraction network learnt on such synthetically distorted videos can be applied to authentically distorted videos to elicit latent quality aware representations. 

The features learnt above are inspired from models based on the natural scene statistics (NSS) of the frames and frame differences of natural videos owing to the design principle involved to obtain quality labels. In addition to the above, we also augment the feature design by concatenating non-NSS based handcrafted features from the successful two-level video quality model (TLVQM) \cite{tlvqm}. The concatenated NSS and motion based non-NSS features capture a wider variety of quality degradation in videos. We show that such a feature combination offers the best performance when trained with limited labels when compared to other features. Thus, they tend to achieve superior performance in SSL. 

In summary, the following are the main contributions of our work:
\begin{enumerate}

\item We generate reliable pseudo-ranks for pairs of unlabelled videos to counter the confirmation bias \cite{mean_teach} due to the noisy pseudo-labels prediction. The reliability of the pseudo-ranks is further enhanced using feedback from a teacher in a student-teacher framework with strong-weak augmentations.  


\item We design a novel strong-weak quality invariant augmentation of videos using temporal subsampling. 

\item We learn frame-level spatial, and temporal quality features from a large corpus of synthetically distorted videos using the spatio-temporal entropic differences index \cite{strred}. This allows rich learning of natural scene statistics based features.


\item We show through detailed experiments that our semi-supervised learning framework achieves the state of the art performance on multiple authentic VQA datasets with limited labels. We also show that our feature learning framework is important in achieving the state of the art performance.  
\end{enumerate}

\section{Related Work}
We survey related work under various categories of NR VQA approaches. 

\textbf{Handcrafted Features for NR VQA}. One of the most successful NR VQA approaches is based on modelling natural video statistics, and extracting features from such models. Statistical models for the discrete cosine transform (DCT) coefficients of frame differences \cite{vbliind}, 3D-DCT coefficients \cite{3d_dct}, three-dimensional mean subtracted contrast normalized coefficient's \cite{nstss}, and optical flow \cite{optical_nrvqa} have been effectively used for NR VQA. 
Several early NR VQA models were developed based on features such as sharpness, noise, blockiness, and temporal correlations \cite{nrvqa_1,nrvqa_2,nrvqa_3}. A recent successful model \cite{tlvqm} employs low complexity features from all the frames, and high complexity features from a subset of frames to effectively model the distortions in consumer generated content. 
The VIDEVAL model \cite{utube_ugc} adopts an approach of feature selection from features of existing different image, and video quality assessment models. 

\textbf{CNN-based NR VQA}. One broad set of approaches that use CNNs involves the use of CNNs in conjunction with other heuristics. The 3D shearlet transform output was processed using CNNs to predict video quality in one of the first attempts in this approach \cite{saconva}. A combination of spatial features from CNNs with handcrafted features for temporal cues was utilized for NR VQA \cite{cnnvqaicip2018}. CNN features have also been combined with heuristic feature based methods to achieve the state of the art NR VQA performance \cite{cnntlvqm,rapique}. 

On the other hand, a few methods design a fully CNN based approach for NR VQA. An end to end deep learning framework was developed to predict compressed video quality for specific codecs \cite{vmeon}. Motion representations have also been derived in an end-to-end manner for NR VQA \cite{rirnet}. The use of 3D CNNs was explored along with long short term memory units \cite{deep3dcnnvqa}. Pre-trained ResNet50 \cite{resnet} features trained for image classification are passed through gated recurrent units for successful NR VQA \cite{qa_in_the_wild}. PVQ \cite{patchVQ} extracts both 2D, and 3D features from pre-trained PaQ-2-PiQ \cite{paq-2-piq}, and 3D ResNet-18 respectively to predict global video quality. MLSP-VQA-FF \cite{konvid150k} extracts features at multiple levels from the pre-trained Inception network and regresses against ground truth quality.  Multiscale end-to-end NR-VQA algorithm \cite{hierarchical_nrvqa} has been proposed that adopts a hierarchical fusion of features at each scale and maps these features onto ground truth quality. FAST-VQA \cite{fastVQA} employs efficient sampling techniques and finetunes a pre-trained Swin-T transformer \cite{swinT} applied on these patches \cite{patchVQ}.

\textbf{Weakly supervised NR VQA}. Zhang et al. \cite{weakly_sup} consider weakly supervised NR VQA where they learn features by predicting a full-reference image quality measure on the frames. The learnt features are then regressed against all the available human opinion scores through a quality score histogram feature. Further, a resampling strategy is also adopted to select appropriate samples. While our approach also uses a full-reference objective model to learn features, we learn both spatial, and temporal features using the perceptually relevant ST-RRED model. The work by Zhang et al. does not consider how such features can be applied to learn quality on authentically distorted videos with limited labels. UCDA \cite{ucda} explores unsupervised domain adaptation from synthetic video distortions to authentic distortions. However, the method requires a large number of human labels in the source domain for its effective performance in the target domain. 

\textbf{Self-supervised feature learning for NR VQA}. The Video CORNIA model \cite{vcornia} adopts a dictionary learning approach to learn frame-level quality features. However, since it relies on training with full-reference video quality measures, it cannot be used for authentically distorted videos. 
CSPT \cite{cspt} is a pre-training based self-supervised learning method which learns quality aware features through the video frame prediction task. The resulting features are used to predict quality using full supervision. VISION \cite{vision} learn spatio-temporal quality aware feature using multiview contrastive learning from unlabelled videos. But, VISION and CSPT do not perform well when evaluated in full supervised evaluation setting.

\textbf{Unsupervised NR VQA}. The VIIDEO \cite{viideo} model represents a completely blind NR VQA model which does not involve training of any kind \cite{viideo}. The model identifies intrinsic statistical regularities in natural videos, and measures deviations in such properties when distortions are introduced. STEM\cite{STEM} and NVQE \cite{NVQE} also adopts a similar approach by combining NIQE \cite{niqe} with the perceptual straightening hypothesis on temporal information. Nevertheless, it may be possible to perform better than the unsupervised models by utilizing labelled, and unlabelled distorted videos through semi-supervised learning.

\section{Methodology}
\subsection{Problem Formulation}
Given a set $\mathcal{V}$ of labelled (or annotated with human opinion subjective scores) videos, and set  $\mathcal{U}$ of unlabelled videos, the goal is to learn an NR VQA model that can predict the video quality without the availability of a reference. The model is evaluated on a test set $\mathcal{W}$. We note that the sets $\mathcal{V}$, $\mathcal{U}$ and $\mathcal{W}$ have some similarity in the nature of distortions that one can see. This work assumes that all these are obtained by splitting a given VQA database containing authentically distorted videos into non-overlapping sets. Note that a reference video is not available for any of the videos in $\mathcal{V}$, $\mathcal{U}$ and $\mathcal{W}$. 

Since the set $\mathcal{V}$ of labelled videos is very small, it may be challenging to learn a model from scratch on such limited data. 
Thus we focus on semi-supervised learning to learn a relevant mapping from quality aware feature representations to the video quality score.
Robust features can give a better baseline when regressed against ground truth than learning a model from scratch on very limited labelled data. 

\subsection{Overview} \label{overview}
We adopt a consistency regularization based student-teacher approach in conjunction with strong-weak quality invariant augmentations. In particular, a student model is trained with the labelled data and strong augmentations of the unlabelled data. The pseudo-labels for the unlabelled data are provided by the teacher model applied on weak augmentations of the unlabelled data. As mentioned in \cite{survey_semiVQA}, consistency regularization based models such as mean-teacher \cite{mean_teach}, noisy-student \cite{noisy_student} etc., suffer from a confirmation bias in the target pseudo-labels due to the noisy predictions. To overcome the above limitation, we compute pseudo-ranks instead of pseudo-labels from the predictions generated by the teacher. We further improve the reliability of the pseudo-ranks by filtering out unlabelled video pairs which fail our threshold criteria. 
We describe each stage in detail in the following.

\subsection{NR VQA Model Architecture} \label{architecture}

In this section, we discuss typical architectures of a quality aware model to estimate video quality. We apply our semi-supervised learning framework on such architectures. Video quality features can be frame level or video level or a hybrid of both types of features. We focus mainly on how pre-trained features can be mapped to perceptual quality using semi-supervised learning. We describe these frameworks as follows. 
 
\subsubsection{Frame Level Feature Model} \label{frame_lvl_arch}
 
\begin{figure}
\centering
\includegraphics[width=\columnwidth]{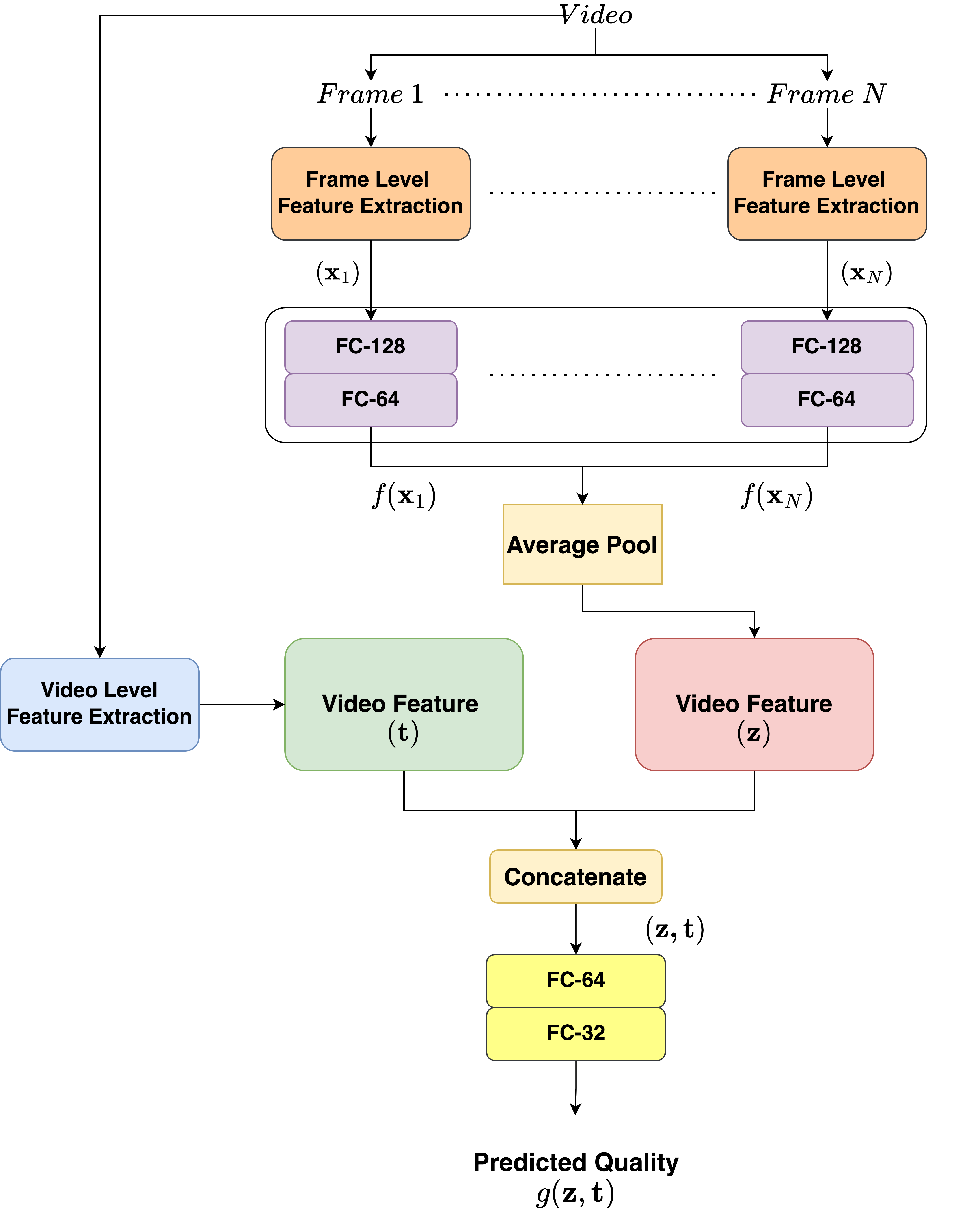}
\caption{ Frame level features from a video are fed to a fully connected network to get video level representation $\mathbf{z}$ as in Equation \ref{frame2vid}. These representations are further concatenated with other video level feature such as $\mathbf{t}$. Hybrid video level representation are then regressed against Mean Opinion Scores using a fully connected network as in Equation \ref{eqn:hybrid_qamodel}.}

\label{fig:framework}
\end{figure}

Let the spatio-temporal features extracted at the frame level of a video $\mathbf{v}$ be $\mathbf{x}_n$, where $n$ is the frame index, $n\in\{1,2,\dots,N\}$, and $N$ is the total number of frames in a given video. Examples of such frame level features include those such as VSFA \cite{qa_in_the_wild} and HEKE\cite{heke-csvt}. As shown in Figure \ref{fig:framework}, the spatio-temporal features are passed through two fully connected layers with ReLU, and sigmoid non-linearity respectively, represented as function $f$, to obtain an output $f(\mathbf{x}_n)$. 
We then temporally average the features across all the $N$ frames to obtain a video level feature $\mathbf{z}$ as
\begin{equation} \label{frame2vid}
  \mathbf{z} = \frac{1}{N}\sum_{n=1}^N f(\mathbf{x}_n). 
\end{equation}
The temporal averaging of features to arrive at a global video feature is more beneficial than passing them through recurrent layers as observed in literature \cite{cnntlvqm}. $\mathbf{z}$ can be further passed through fully connected layers to output a scalar that represents the video quality. Hence we learn a mapping from $\mathbf{z}$ to perceptual quality using fully connected layers $g(\cdot)$ as 

\begin{equation}\label{eqn:frame_qamodel}
  \hat{q}(\mathbf{v})=g(\mathbf{z}).
\end{equation}
The goal of semi-supervised learning of frame level features is the learn the mappings $f(\cdot)$ and $g(\cdot)$. 
 
\subsubsection{Video Level Feature Model} \label{video_lvl_arch}

Handcrafted or CNN based video level features denoted as $\mathbf{t}$ is passed through a fully connected network $g(\cdot)$ as above to estimate video quality as 

\begin{equation}\label{eqn:video_qamodel}
  \hat{q}(\mathbf{v})=g(\mathbf{t}).
\end{equation}
Examples of such video level features include TLVQM \cite{tlvqm}, and VIDEVAL \cite{videval}. Note that these features are obtained by taking the video as a whole and not obtained by averaging frame level features. In this scenario, we only learn the mapping $g(\cdot)$ through semi-supervised learning. 

\subsubsection{Hybrid Feature Model} \label{hybrid_lvl_arch}

When both frame and video level features are present, we combine the above two models. In general, we concatenate handcrafted, or CNN based video level features denoted as $\mathbf{t}$ with $\mathbf{z}$ to estimate the overall video quality. The pair $(\mathbf{z},\mathbf{t})$ is then fed to $g(\cdot)$ to predict the perceptual quality of the video as,

\begin{equation}\label{eqn:hybrid_qamodel}
  \hat{q}(\mathbf{v})=g(\mathbf{z},\mathbf{t}).
\end{equation}
In this setup, both $f(\cdot)$ and $g(\cdot)$ are learnt during semi-supervised learning. 


\subsection{Semi-supervised Learning by Generating Pseudo-Rank Pairs}\label{sec:pairwise}

The core idea of our approach is to learn $f(\cdot)$ and $g(\cdot)$ or $g(\cdot)$ alone is to enforce consistency of pseudo-ranks for unlabelled video pairs. In this regard, we consider a student-teacher based consistency regularization approach, where two models are maintained, a student model and a teacher model. While the student model is trained with both labelled and augmented unlabelled data, the teacher model provides weak supervision for the student model on the augmented unlabelled data to improve its performance. Initially, both the models are identical but as learning progresses, the teacher model is obtained as a weighted average of the network parameters of the past teacher model and the current student model. Further, we deploy a novel quality invariant strong-weak augmentation strategy while training the student and teacher models. Existing augmentations in literature such as noisy input or photometric transformed  data \cite{survey_semiVQA} may not be appropriate for VQA task since they can alter the quality of the videos.


In recent VQA works such as TLVQM \cite{tlvqm}, VISION \cite{vision}, and VIDEVAL \cite{videval}, it has been shown that the quality estimated from video frames sub-sampled to as low as 1 frame per second is approximately equivalent to the quality of the video at the full frame rate. Thus, we design an augmentation based on video subsampling of the frames. For the unlabelled data, the frame-level features temporally sub-sampled at 1 frame per second (referred to as \emph{strong augmentation}) are input to the student model. The target pseudo-label is provided by the teacher model, which makes a prediction based on the frame-level features, temporally sub-sampled at half the original frame rate (referred to as \emph{weak augmentation}).  
Note that the above augmentation strategy can be applied to our student-teacher based model only when the quality aware representations considered are available at a frame level.

To counter confirmation bias in the teacher model prediction on the weakly augmented videos, we generate pseudo-ranks instead of pseudo-labels for a pair of unlabelled videos. Our hypothesis is that if the teacher model quality predictions of two unlabelled videos $\mathbf{u}_1$, and $\mathbf{u}_2$ differ by greater than a threshold $\tau$, then the pairwise quality ranking of the videos inferred from their predicted qualities is likely to be correct. Thus, we generate pairwise pseudo-ranks of the unlabelled videos in terms of their qualities predicted by the teacher model, and use these ranks to supervise the student model.

Mathematically, we create two models $q_s$, and $q_t$ corresponding to the student and teacher models similar to Equation (\ref{eqn:hybrid_qamodel}). Initially, both models are identical. 
For any video $\mathbf{u}$, its quality prediction using the student, and teacher models are obtained as $q_s(\mathbf{u})$, and $q_t(\mathbf{u})$ while the corresponding ground truth is denoted as $q(\mathbf{u})$. If video $\mathbf{u}$ has a frame-rate $r$, let $\mathcal{T}_s(\mathbf{u})$, and $\mathcal{T}_w(\mathbf{u})$ be the strong and weak augmentation functions which select frames at frame rates of 1 fps and $r/2$ fps respectively.  For a pair of videos $\mathbf{u}_1$, and $\mathbf{u}_2$, with $\mathbf{u}_1,\mathbf{u}_2\in\mathcal{U}$, and $|q_t(\mathcal{T}_w(\mathbf{u}_1))-q_t(\mathcal{T}_w(\mathbf{u}_2))|>\tau$, we define the pairwise pseudo-ranking as
\begin{equation}
r(\mathbf{u}_1,\mathbf{u}_2) =
  \begin{cases}
   1 & \text{if $q_t(\mathcal{T}_w(\mathbf{u}_1))\geq q_t(\mathcal{T}_w(\mathbf{u}_2))$}\\
   0 & \text{otherwise}
  \end{cases}    .
\end{equation}
\par
We update the student model's parameters using the available labels, and the pseudo-rank pairs. In particular, the student model is trained to ensure that it satisfies the ranking of videos according to the pseudo-rank pairs generated by the teacher model. Enforcing the student model prediction to match the pairwise pseudo-ranks of videos generated by the teacher model achieves consistency regularization by being invariant to the different augmentations that are applied as input to student-teacher models. 


\begin{figure*}
\centering
\includegraphics[width=\textwidth]{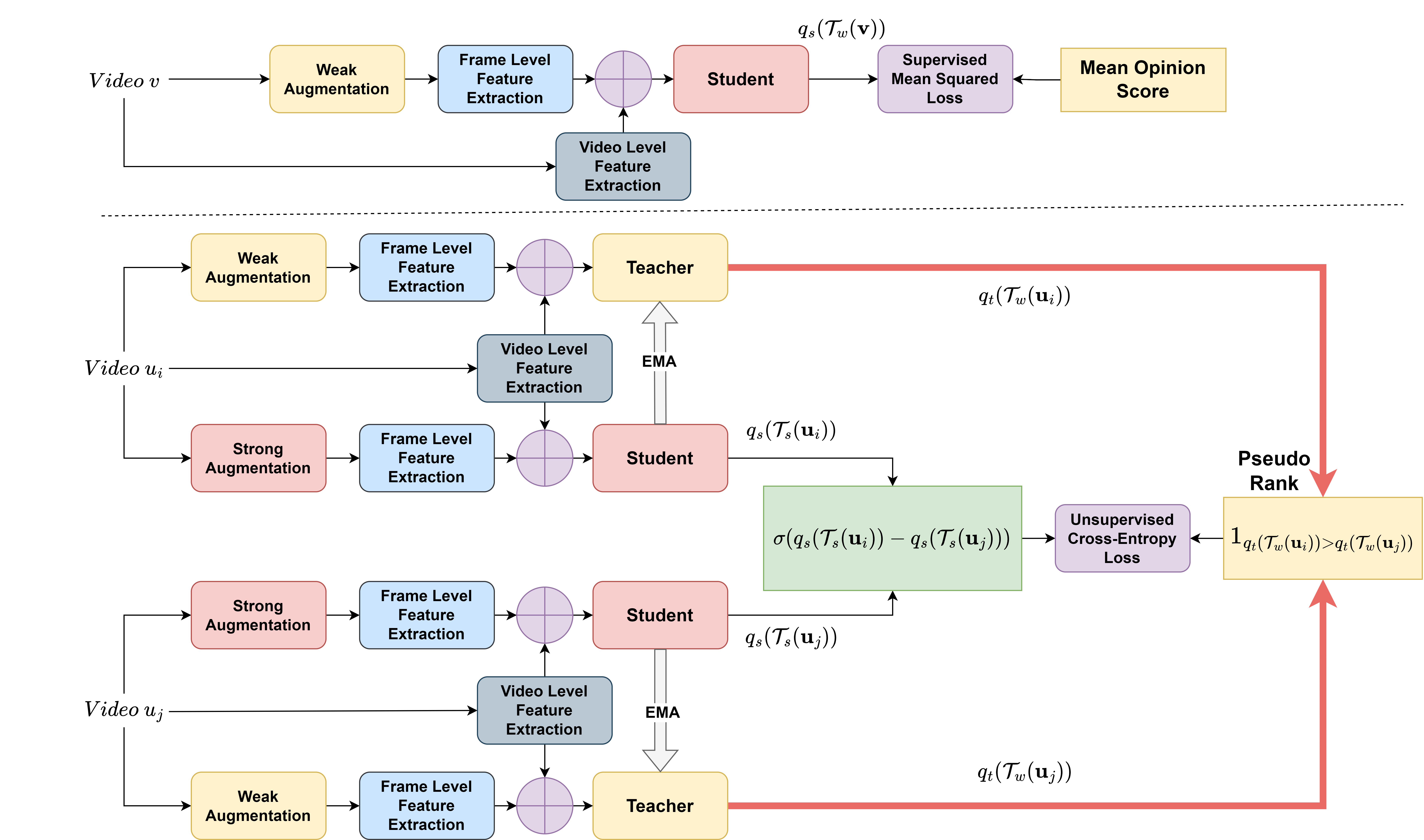}
\caption{Framework of our semi-supervised learning approach on authentically distorted camera captured videos. $\mathbf{v}$ is a video belonging to the labelled set $\mathcal{V}$, and $\mathbf{u}_i$ and $\mathbf{u}_j$ are a pair of videos belonging to the unlabelled set $\mathcal{U}$. Note that all the student models weights are shared.}

\label{fig:lpr}
\end{figure*}

Using the generated pairwise pseudo-ranks, we deploy rank based learning that has been widely studied in the literature \cite{rankNet,dipIQ}, where a Siamese network is used to predict the target from a pair of data sample. Note that our main contribution is in the generation of reliable pseudo-ranks and once the pseudo-ranks are generated, we use rank based learning methods as studied in the literature. We use the quality predictions ${q}_s(\mathcal{T}_s(\mathbf{u}_1))$, and $q_s(\mathcal{T}_s(\mathbf{u}_2))$ of videos $\mathbf{u}_1$, and $\mathbf{u}_2$ satisfying $|q_t(\mathcal{T}_w(\mathbf{u}_1))-q_t(\mathcal{T}_w(\mathbf{u}_2))|>\tau$ to compute the probability that $q_s(\mathcal{T}_s(\mathbf{u}_1))>q_s(\mathcal{T}_s(\mathbf{u}_2))$ as
\begin{align}
  \hat{r}(\mathbf{u}_1,\mathbf{u}_2) &= \sigma(q_s(\mathcal{T}_s(\mathbf{u}_1))-q_s(\mathcal{T}_s(\mathbf{u}_2))) \\
  &= \frac{\exp(q_s(\mathcal{T}_s(\mathbf{u}_1))-q_s(\mathcal{T}_s(\mathbf{u}_2)))}{1+\exp(q_s(\mathcal{T}_s(\mathbf{u}_1))-q_s(\mathcal{T}_s(\mathbf{u}_2)))}.
\end{align}

Thus the unsupervised loss for training the student model on unlabelled set $\mathcal{U}$ is given as,
\begin{equation} \label{eqn:unsupervised_loss}
    \mathcal{L}_u = \sum_{\substack{(\mathbf{u}_1,\mathbf{u}_2) \in \mathcal{U}\\|q_t(\mathcal{T}_w(\mathbf{u}_1))-q_t(\mathcal{T}_w(\mathbf{u}_2))|>\tau}}\mathcal{L}_{cross}(r(\mathbf{u}_1,\mathbf{u}_2), \hat{r}(\mathbf{u}_1,\mathbf{u}_2))
\end{equation}
where, $\mathcal{L}_{cross}(p_1,p_2)$ is the binary cross entropy defined as 
\begin{equation}
  \mathcal{L}_{cross}(p_1,p_2) = -p_1\log p_2-(1-p_1)\log(1-p_2).
\end{equation}
For every video $\mathbf{v}$ in the labelled set $\mathcal{V}$, its quality prediction using the student model is obtained as $q_s(\mathcal{T}_w(\mathbf{v}))$ and the corresponding ground truth is denoted as $q(\mathbf{v})$. The supervised loss on the labelled set is given as
\begin{equation} \label{eqn:supervised_loss}
    \mathcal{L}_s = \sum_{v\in\mathcal{V}}|q(\mathbf{v})-q_s(\mathcal{T}_w(\mathbf{v}))|.
\end{equation}
The overall objective function for training the student network is given as

\begin{equation} \label{eqn:loss}
    \mathcal{L} = \mathcal{L}_s + \lambda \mathcal{L}_u,
\end{equation}

where $\lambda$ represents the relative weight between the two losses. $\lambda$ is chosen such that the order of magnitude of the supervised and unsupervised loss terms are similar so that the unsupervised loss cannot overpower the effect of supervised loss.

Suppose the parameters of the teacher model and student model at iteration $n$ are given by $\theta_s^{(n)}$ and $\theta_t^{(n)}$. The teacher model is then updated as the moving average of consecutive student model similar to Mean Teacher \cite{mean_teach}
\begin{equation}
  \theta_t^{(n)} = \alpha \theta_t^{(n-1)}+(1-\alpha)\theta_s^{(n)}.
  \label{eqn:mean_teacher}
\end{equation}
In principle, since the teacher model is updated at every iteration, the pseudo-rank pairs must be updated every iteration. To limit the computational overhead of generating the rank pairs of all the unlabelled videos at every iteration, we update the pseudo ranks after every $K$ training iterations, although the teacher model is updated every iteration. We refer to our entire learning framework consisting of student-teacher models, strong-weak augmentations and pseudo-rank generation as Learning with Pseudo Ranks (LPR).

\section{Quality Aware Feature Representations} \label{sted_feature}

The semi-supervised learning approach proposed in the previous section is built on top of quality aware feature representations, where the focus is to learn the regression models to predict quality. In VQA literature, features can be mostly subdivided as natural scenes statistic (NSS) based and non-NSS based. NSS features are built on the statistical regularities observed in intensity, colour, spatio-temporal frequencies, spatial correlation among pixels, and so on. Any deviation in NSS can be used to estimate quality degradation in videos. Non-NSS based features involve sharpness, camera shake, and also object motion related distortions \cite{tlvqm}. 

While NSS based features exist in the literature for images and videos, here we present a particular approach to capture NSS-based features of videos using transfer learning. In particular, we build on the robust performance of the spatio-temporal entropic differences (ST-RRED) \cite{robust_strred} index for compression and transmission distortions. ST-RRED is a natural scene statistics based approach that computes the localized entropic differences between the reference and distorted video frames and frame differences. The spatio-temporal entropic differences were recently predicted in a no-reference manner and its utility in achieving robust generalization performance in measuring compression and transmission distortions was shown in \cite{nrsted}. In this work, we learn CNN based features that can predict the spatial reduced reference entropic differences (SRRED) and temporal reduced reference entropic differences (TRRED) from frames and frame differences respectively on synthetically distorted (compression, transmission and noise) videos. Although these features are learnt to predict SRRED and TRRED for synthetic distortions, they contain some latent representations of video quality that can be leveraged for predicting the quality of authentically distorted videos. 

\begin{figure*}
\centering
\includegraphics[width=0.8\textwidth]{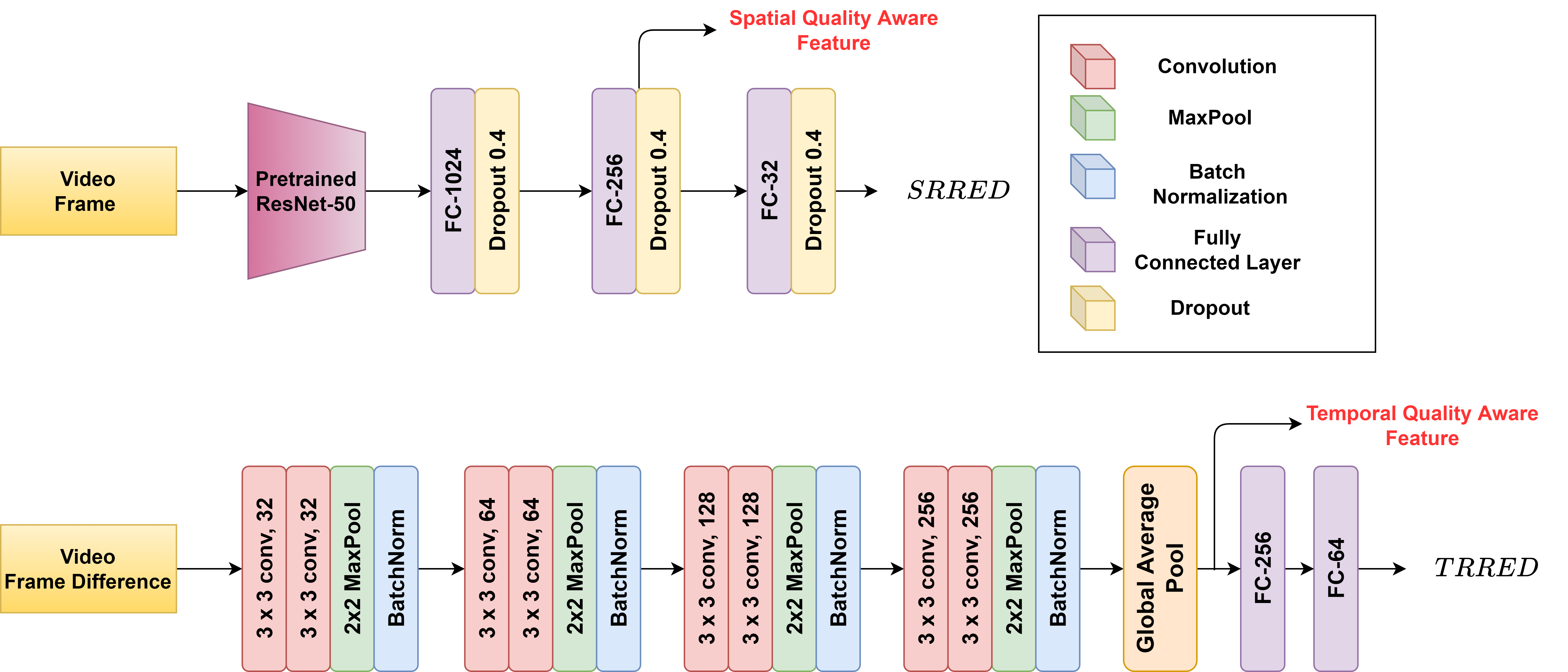}
\caption{The overall structure of STED feature learning on synthetically distorted videos. Spatial quality features are learnt by training a network to predict SRRED indexes from video frames. Similarly, temporal features are learnt from frame differences by regressing against TRRED indexes. Also mentioned is the layer at which spatial and temporal features are extracted.}
\label{fig:feature_learning}
\end{figure*}

We deploy a pair of CNNs to learn SRRED and TRRED at a frame level respectively as shown in Figure \ref{fig:feature_learning}. A pre-trained ResNet-50 architecture followed by three fully connected layers with ReLU nonlinearity is deployed to learn SRRED from video frames. 
Note that the output of ResNet-50 is globally spatial average pooled to arrive at a 2048 dimensional vector irrespective of the spatial resolution of the input video frame. 
Several pieces of literature have shown that pre-trained ResNet-50 features trained for image classification can be effectively transferred for image quality assessment \cite{bovik_deep_feat, unreason_deepfeat}. In a similar vein, the NR-STED framework also showed \cite{nrsted} that pre-trained ResNet-50 features can effectively be used to predict SRRED. We use the L2 loss between the predicted SRRED and the ground truth SRRED to train the fully connected layers. We tap the 256 dimensional feature obtained in the penultimate layer as the spatial feature extractor for authentically distorted videos. 

To predict TRRED from frame differences, we use a simpler network shown in Figure \ref{fig:feature_learning} and train it from scratch. 
The network used here is simpler than the one used in SRRED prediction as we find that it trains faster without compromising on the prediction performance. 
We use the L1 loss between the predicted TRRED and the ground truth TRRED to train all the layers. The motivation to use the L1 loss stems from the robustness to outliers when training with a small number of samples in a batch owing to memory constraints. We tap the 256 dimensional vector after the global average pooling layer as the temporal feature extractor for authentically distorted videos. Video quality features extracted from our learned model are referred to as spatio-temporal entropic difference (STED) features. STED features are extracted at a frame level across all frames in a video. The resulting features extracted at a frame level using SRRED or TRRED are concatenated and this concatenated vector corresponds to $\mathbf{x}_n$ in our frame level feature model described in Section \ref{frame_lvl_arch}.

The idea of using frame level objective scores for training CNNs has been explored in literature \cite{weakly_sup}. However, only image quality measures have been used and existing approaches do not effectively capture the temporal distortions. The use of SRRED and TRRED enables the effective modelling of spatial and temporal distortions at the frame level. Although the learning of full reference measures such as ST-RRED exists in  literature \cite{nrsted, heke-csvt}, the novelty of our work is in using the latent features learnt during this process for VQA of authentically distorted videos.  

The above feature learning method is implicitly based on an NSS approach owing to the use of SRRED, and TRRED during training. TLVQM \cite{tlvqm} addresses the potential inefficiencies in NSS, and designs heuristic features to capture blockiness, sharpness extremes, and camera shake. Thus, the features in the TLVQM \cite{tlvqm} approach are complementary to the NSS based STED features. 
While STED features are computed at a frame level using our pre-trained model above, TLVQM features are video level features. Thus, in the hybrid NR VQA model in Section \ref{hybrid_lvl_arch}, STED features $\mathbf{x}_n$ are fed as frame level quality features while TLVQM features $\mathbf{t}$ are concatenated with video level STED features at a later stage as in Equation \ref{eqn:hybrid_qamodel}. While CNN-TLVQM \cite{cnntlvqm} also uses image quality based CNN features to boost the performance of TLVQM, the features we learn based on ST-RRED are richer than the CNN features in \cite{cnntlvqm} owing to the modelling of both spatial, and temporal distortions. Thus, there is scope for combining the STED features we learn with the TLVQM features to obtain a richer set of features for learning video quality with limited labels. In Section \ref{QA_feature},  we analyze the performance of the STED-TLVQM features and show that it achieves better performance than other CNN based and handcrafted features based methods in the low data regime.

\section{Experiments and Results}\label{sec:experiments}

\subsection{Databases}

We evaluate our semi-supervised video quality learning method on three popular authentically distorted VQA datasets described as follows:

\subsubsection{KoNViD-1K \cite{konvid}}
This dataset contains 1200 videos with a wide variety of content, distortion types and subjective quality variations. The videos are of $960 \times 540$ resolution, correspond to a frame rate of 24, 25 or 30 frames per second, and are of 8 seconds in duration. 

\subsubsection{LIVE Video Quality Challenge (VQC) Database \cite{livevqc}} The LIVE VQC database consists of 585 videos of unique content available at 18 different  spatial resolutions ranging between $1980\times 1080$ to $320\times 240$ across landscape and portrait modes. 
All the videos are 10 seconds long. 

\subsubsection{LIVE Qualcomm Database \cite{liveqcomm}} This database consists of 208 videos accounting for distortions generated during the camera capture process using eight mobile devices. The videos are of spatial resolution $1920\times 1080$, 15 seconds long when played at 30 fps. 

Similar to \cite{cnntlvqm}, since our focus is on authentically distorted videos through camera capture, we omit the YouTube UGC dataset \cite{utube_ugc}, since it  contains a large fraction of artificially generated content in the form of animations and computer graphics. 

For learning STED features as described in Section \ref{sted_feature}, we use several synthetic databases such as the LIVE Mobile VQA dataset \cite{mobile1}, LIVE VQA dataset \cite{live_sd1}, EPFL-Polimi dataset \cite{epfl1}, ECVQ and EVVQ datasets \cite{ecvq_evvq1} and the CSIQ database \cite{csiq}.  In particular, we only used the videos from these synthetic datasets and do not use any subjective scores. The features are learned to predict the SRRED and TRRED on these videos since a reference video is available in all these synthetic datasets.

\subsection{Experimental Setting}
Semi-supervised methods are typically evaluated by treating most of the dataset as unlabelled and using a small part of the dataset as labelled. We first divide the dataset into training and testing in the ratio of 80\% and 20\%. We evaluate the performance when only 30, 60 and 120 videos belonging to the training set are labelled in the form of mean opinion scores. Further, the videos with labels are randomly sampled from the training set. We conduct our experiments on ten different splits of the dataset into training and testing and report the median performance. 

We evaluate the performance of VQA methods using the conventional measures such as Spearman's rank order correlation coefficient (SROCC), Pearson linear correlation coefficient (PLCC) between the predicted quality scores and the ground truth quality scores. 

\subsection{Performance Analysis of Quality Features in Limited labelled Data Regime} \label{QA_feature}
We first conduct an experiment where we compare different video quality features using the limited labelled data and supervised learning. Thus, the unlabelled data is not used during the training in this experiment. The goal of this analysis is to identify features that perform best in the limited labelled data regime. We believe that features that work well in this regime can be bootstrapped to improve performance best with semi-supervised learning. 

\subsubsection{Benchmarking Quality Aware Features}

We compare various CNN and heuristics based feature learning methods designed for quality analysis for limited labels. In particular, we compare STED-TLVQM features described in Section \ref{sted_feature} with classical methods such as Video BLIINDS \cite{vbliind} and Video Cornia \cite{vcornia} features. We find that learning a CNN from scratch on limited data gives poor performance. Thus, we focus on recent heuristic and pre-trained CNN based methods such as TLVQM \cite{tlvqm}, which comprises of motion-based features predominantly, VIDEVAL \cite{videval}, which is a combination of various classical VQA and image QA feature based methods \cite{tlvqm,brisque, friquee}, and VSFA \cite{qa_in_the_wild} which has a pre-trained ResNet50 \cite{resnet} backbone. We also compare with combinations of pre-trained CNN based features with heuristics features such as RAPIQUE \cite{rapique}, and CNN-TLVQM \cite{cnntlvqm}. We evaluate the performance for these features by regressing them using the frame, video, or hybrid model as appropriate using only the labelled videos. 


\begin{table*}
\caption{ SROCC performance analysis and comparison on KoNVid-1K, LIVE VQC, and LIVE Qualcomm datasets. The quality aware feature extraction algorithms are trained on 30, 60, and 120 labelled data respectively.}
\centering
\label{tab:feat_srocc}
\begin{adjustbox}{max width=\textwidth}
\begin{tabular}{|c|c|c|c|c|c|c|c|c|c|}
\hline
& \multicolumn{3}{c|}{KoNVid-1K} & \multicolumn{3}{c|}{LIVE VQC} & \multicolumn{3}{c|}{LIVE Qualcomm} \\
\hline
Algorithm & 30 labels & 60 labels & 120 labels & 30 labels & 60 labels & 120 labels & 30 labels & 60 labels & 120 labels \\
\hline
Video BLIINDS &0.216 &0.364 &0.389 &0.363 &0.512 &0.540 &0.268 &0.411 &0.478\\
VSFA &0.514 & 0.563 &0.636 & 0.505 & 0.560 & 0.583 &0.315 & 0.527 &0.644\\
TLVQM &0.491 &0.576 &0.636 &0.541 & 0.587 & 0.610 &0.417 &0.544 &0.713\\
VIDEVAL &0.463 & 0.520 & 0.593 & 0.533 & 0.573 & 0.614 & 0.402 & 0.492 & 0.571 \\
RAPIQUE &0.498 & 0.570 & 0.635 & 0.541 & 0.611 & 0.637 &0.371 & 0.487 & 0.584\\
HEKE  & 0.463 & 0.504 & 0.566 & 0.438 & 0.490 & 0.569 & 0.375 & 0.496 & 0.583 \\
CNN-TLVQM &0.539 &0.632 &0.653 &0.533 & 0.596 & 0.618 &0.347 &0.543 & 0.655 \\
STED &0.573 &0.634 &0.655 &0.511 &0.556 &0.586 &0.423 &0.530 &0.683\\
\hline
STED-TLVQM & \textbf{0.616} & \textbf{0.665} & \textbf{0.697} & \textbf{0.561} & \textbf{0.658} & \textbf{0.678} & \textbf{0.475} & \textbf{0.591} & \textbf{0.765}\\
\hline
\end{tabular}
\end{adjustbox}
\end{table*}

\begin{table*}
\caption{ PLCC performance analysis and comparison on KoNVid-1K, LIVE VQC, and LIVE Qualcomm datasets. Algorithms and learning methods are similar to that in Table \ref{tab:feat_srocc}.}
\centering
\label{tab:feat_lcc}
\begin{adjustbox}{max width=\textwidth}
\begin{tabular}{|c|c|c|c|c|c|c|c|c|c|}
\hline
 & \multicolumn{3}{c|}{KoNVid-1K} & \multicolumn{3}{c|}{LIVE VQC} & \multicolumn{3}{c|}{LIVE Qualcomm} \\
\hline
Algorithm & 30 labels & 60 labels & 120 labels & 30 labels & 60 labels & 120 labels & 30 labels & 60 labels & 120 labels \\
\hline
Video BLIINDS &0.215 &0.344 &0.372 &0.395 &0.495 &0.551 &0.299 &0.414 &0.525\\
VSFA &0.534 &0.589 &0.640 & 0.553 & 0.612 & 0.644 &0.317 &0.592 &0.661\\
TLVQM &0.501 &0.570 & 0.628 & 0.574 & 0.606 &0.655 &0.443 &0.568 &0.745\\
VIDEVAL & 0.466 & 0.524 & 0.592 & 0.534 & 0.583 & 0.620 & 0.431 & 0.535 & 0.585 \\
RAPIQUE & 0.509 & 0.578 & 0.649 & 0.556 & 0.631 & 0.666 & 0.394 & 0.529 & 0.605\\
HEKE & 0.464 & 0.507 & 0.564 & 0.470 & 0.526 & 0.599 & 0.372 & 0.477 & 0.587 \\
CNN-TLVQM &0.567 &0.635 &0.656 & 0.559 &0.605 &0.648 & 0.405 & 0.559 & 0.680\\
STED &0.598 &0.657 &0.677 &0.519 &0.559 &0.624 &0.462 &0.585 &0.709\\ \hline
STED-TLVQM & \textbf{0.637} & \textbf{0.661} & \textbf{0.707} & \textbf{0.586} & \textbf{0.666} & \textbf{0.692} & \textbf{0.497} & \textbf{0.627} & \textbf{0.763}\\
\hline
\end{tabular}
\end{adjustbox}
\end{table*}

\subsubsection{Training Details}
While STED is trained on synthetically distorted videos, the learned features are then used along with handcrafted TLVQM features in STED-TLVQM. In STED, we train the spatial feature extraction network using SRRED for 20 epochs with a batch size of 16, and Adam \cite{adam} optimizer. Since we train the temporal network from scratch on the synthetic videos, and the video frames at original resolution are fed as input, a batch size of 8 is chosen to train this network using TRRED for 30000 iterations. Note that our spatial, and temporal feature learning framework allows us to train with videos of any resolution. The trained STED model is then used to extract quality aware features from authentically distorted videos. The frame level spatio-temporal features are transformed to video level STED features as in Equation \ref{frame2vid}. The TLVQM features are normalized to lie in the range of $0-1$ by taking into account the minimum and maximum values across dimensions. 

We normalize the mean opinion score (MOS) for the videos to a 0 to 1 scale during training. The network corresponding to parameters of $f(\cdot)$, and $g(\cdot)$ is trained for 1000 iterations with a batch size of 32 for the 60, and 120 labels cases, and 16 for the 30 labels case. Stochastic Gradient Descent (SGD) with an initial learning rate of $10^-1$ and decay rate of $10^-2$ and momentum as $0.9$ is used to train this model. Note that neither the augmentations nor the student-teacher models are necessary in this experiment and only one set of parameters is trained. 
We note that both $f(\cdot)$, and $g(\cdot)$ are trained for frame level features such as HEKE and VSFA or hybrid features (both frame-video level) features such as STED-TLVQM and CNN-TLVQM. For video level features such as TLVQM, RAPIQUE, VIDEVAL and Video BLIINDS, $f(\cdot)$ does not exist. Note that for VSFA, we take the implicit pre-trained ResNet-50 features as the input to our framework.  


\subsubsection{Performance Comparisons}

A comparative study of different features when trained in the limited labelled data regime is presented in Tables \ref{tab:feat_srocc} and \ref{tab:feat_lcc}. 
We find that STED-TLVQM consistently achieves better performance than all the other models in the supervised scenario on all the datasets.   

\subsection{Performance Analaysis of Semi-supervised Learning for VQA}
We now conduct experiments to validate the main contributions of the paper. Since the STED-TLVQM features perform best, we compare different semi-supervised learning frameworks for these features. We show that our semi-supervised learning framework performs better than other frameworks. 

\begin{table*}
\caption{ SROCC performance analysis and comparison of semi-supervised algorithms applied on STED-TLVQM features on KoNVid-1K, LIVE VQC, and LIVE Qualcomm datasets. The semi-supervised algorithms are Pseudo-Label (PS), Mean Teacher (MT), Noisy Student (NS), $FixMatch^*\ (FM^*)$ with our augmentation and Learning Pseudo-Rank (LPR). The baseline performance is that of STED-TLVQM features trained with 30, 60, and 120 labels.}
\centering
\label{tab:srocc}
\begin{adjustbox}{max width=\textwidth}
\begin{tabular}{|l|c|c|c|c|c|c|c|c|c|}
\hline
& \multicolumn{3}{c|}{KoNVid-1K} & \multicolumn{3}{c|}{LIVE VQC} & \multicolumn{3}{c|}{LIVE Qualcomm} \\
\hline
Algorithm & 30 labels & 60 labels & 120 labels & 30 labels & 60 labels & 120 labels & 30 labels & 60 labels & 120 labels \\
\hline

baseline &0.616 &0.665 &0.695 &0.561 &0.658 &0.678 &0.475 &0.591 &0.765\\ \hline
 \quad+ PL  &0.620 & 0.671 & 0.698 & 0.565 & 0.661 & 0.689 & 0.478 & 0.602 & 0.774 \\
 \quad+ MT &0.623 & 0.674 & 0.699 & 0.569 & 0.666 & 0.691 & 0.481 & 0.616 &0.769\\
\quad+ NS & 0.625 & 0.675 & 0.700 & 0.569 & 0.662 & 0.686 & 0.509 & 0.604 & 0.775 \\
\quad+ $FM^*$ &0.629 & 0.674 & 0.698 & 0.570 & 0.661 &0.702 & 0.503 & 0.609 & 0.773\\ 
\hline
 \quad+ LPR & \textbf{0.675} & \textbf{0.708} & \textbf{0.750} & \textbf{0.621} & \textbf{0.709} & \textbf{0.751} & \textbf{0.557} & \textbf{0.664} & \textbf{0.794}\\

\hline
\end{tabular}
\end{adjustbox}
\end{table*}

\begin{table*}
\caption{ PLCC performance analysis and comparison on KoNVid-1K, LIVE VQC, and LIVE Qualcomm datasets. The semi-supervised algorithms are Pseudo-Label (PS), Mean Teacher (MT), Noisy Student (NS), $FixMatch^*\ (FM^*)$ with our augmentation, and Learning Pseudo-Rank (LPR). The baseline performance is that of STED-TLVQM features trained with 30, 60, and 120 labels.}
\centering
\label{tab:lcc}
\begin{adjustbox}{max width=\textwidth}
\begin{tabular}{|l|c|c|c|c|c|c|c|c|c|}
\hline
& \multicolumn{3}{c|}{KoNVid-1K} & \multicolumn{3}{c|}{LIVE VQC} & \multicolumn{3}{c|}{LIVE Qualcomm} \\
\hline
Algorithm & 30 labels & 60 labels & 120 labels & 30 labels & 60 labels & 120 labels & 30 labels & 60 labels & 120 labels \\
\hline

baseline &0.627 &0.661 &0.705 &0.586 &0.666 &0.692 &0.497 &0.627 &0.763\\ \hline
 \quad+ PL  & 0.631 & 0.669 & 0.708 & 0.585 & 0.667 & 0.695 & 0.517 & 0.631 & 0.773\\
 \quad+ MT & 0.632 & 0.676 & 0.708 & 0.596 & 0.669 & 0.713 & 0.520 & 0.639 & 0.773\\
 \quad+ NS & 0.636 & 0.675 & 0.707 & 0.595 & 0.668 & 0.700 & 0.549 & 0.627 & 0.778\\
 \quad+ $FM^*$ &0.640 & 0.677 & 0.707 & 0.598 & 0.667 & 0.720 & 0.548 & 0.640 &0.774\\ 
\hline
 \quad+ LPR & \textbf{0.668} & \textbf{0.711} & \textbf{0.749} & \textbf{0.615} & \textbf{0.694} & \textbf{0.762} & \textbf{0.572} & \textbf{0.684} & \textbf{0.799}\\

\hline
\end{tabular}
\end{adjustbox}
\end{table*}

\subsubsection{Benchmarking Other Semi-supervised Algorithms} 

The problem of semi-supervised VQA has not been studied much in the literature to the best of our knowledge. Thus there are no standard benchmarks available for comparison. Further, methods typically used in image/video classification literature based on pseudo-labelling, and data augmentation do not easily extend to the VQA problem. In particular, data augmentation strategies that modify the video frames by adding noise or changing brightness levels or contrast can end up modifying the video quality itself, and therefore are not appropriate for VQA. Pseudo-labelling approaches in semi-supervised learning for classification use the label with the maximum confidence as the pseudo-label for the unlabelled data. These types of methods are not suitable for regression tasks such as VQA. Nevertheless, we compare our LPR method with Pseudo-label (PL) \cite{pseudolabel} based methods, and student-teacher based methods such as Mean Teacher (MT) \cite{mean_teach}, and Noisy Student (NS) \cite{noisy_student}. While Mean Teacher uses a teacher model to give pseudo-labels for unlabelled samples, Noisy Student follows a knowledge distillation strategy where the student learned on both labelled and unlabelled samples becomes the new teacher to generate pseudo-labels. We note that, any photometric augmentation used in the above methods for the VQA task was removed. FixMatch \cite{fixmatch} uses photometric transformation based strong-weak augmentation strategy on student-teacher based model. Thus we replace FixMatch strong-weak augmentation without quality invariant strong-weak augmentation. In the rest of the paper we will address this modified FixMatch as $ FixMatch^*$.

\subsubsection{Semi-supervised Training Details}

Initially, the network parameters corresponding to $f(\cdot)$ and $g(\cdot)$ are trained for 1000 iterations using just the supervised loss as in Equation \ref{eqn:supervised_loss} and SGD with an initial learning rate of $10^-1$ and decay rate of $10^-2$ and momentum as $0.9$. 
We then incorporate the pairwise pseudo-rank based loss to fine-tune the parameters of the student model corresponding to $f(\cdot)$, and $g(\cdot)$ with $\lambda=0.1$. These parameters are trained for 1000 iterations with a pseudo-rank update for the unlabelled data after every $K=50$ iterations. The teacher model used to update the pseudo-ranks has a smoothing co-efficient $\alpha = 0.5$ referred to in Equation (\ref{eqn:mean_teacher}). As in the previous stage, we employ a batch size of 32 for the 60, and 120 labels cases, and 16 for the 30 labels case. Since the predicted quality score lies in the 0 to 1 scale, we choose a threshold of $\tau=0.1$ to select pairs of videos with pseudo-rank labels. 

\subsubsection{Performance Comparisons}

We compare the performance of our method (LPR) against Pseudo-labelling (PL), Mean Teacher (MT), Noisy Student (NS), and $ FixMatch^*$ on KoNVid-1K, LIVE VQC, and LIVE Qualcomm datasets in Tables \ref{tab:srocc} and \ref{tab:lcc}. 
We see that LPR not only outperforms other semi-supervised methods on the three authentically distorted databases but also shows considerable improvement over the baseline supervised model trained on low data.

We also conduct statistical significance tests to validate the importance of the correlation coefficient differences observed in Tables \ref{tab:srocc} and \ref{tab:lcc}. These results are given in the supplementary.  

\begin{figure*}
     \centering
     \begin{subfigure}[b]{0.32\textwidth}
         \centering
         \includegraphics[width=\textwidth]{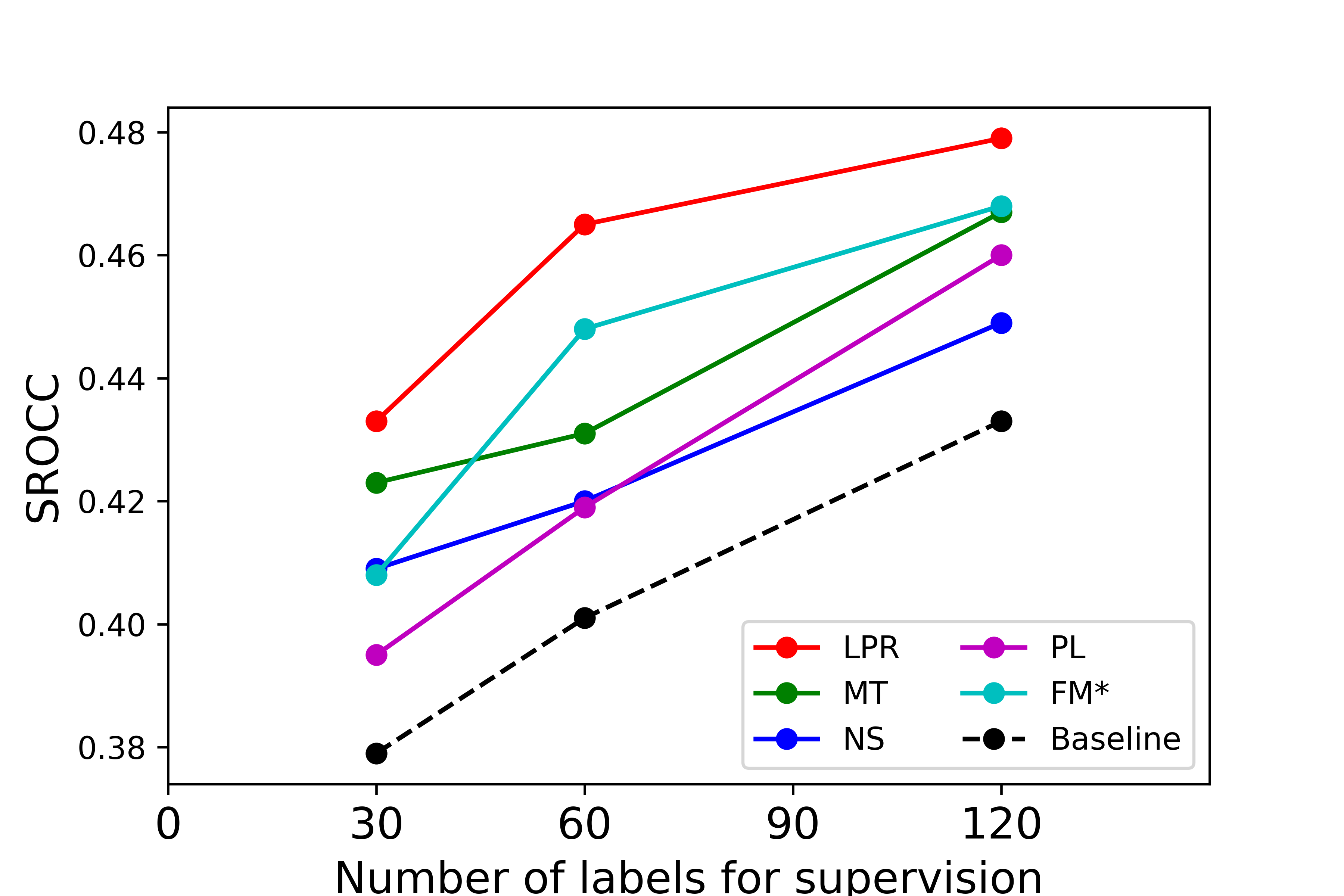}
         \caption{\footnotesize Train: KoNVid-1K; Test: LIVE VQC }
     \end{subfigure}
     \hfill
     \begin{subfigure}[b]{0.32\textwidth}
         \centering
         \includegraphics[width=\textwidth]{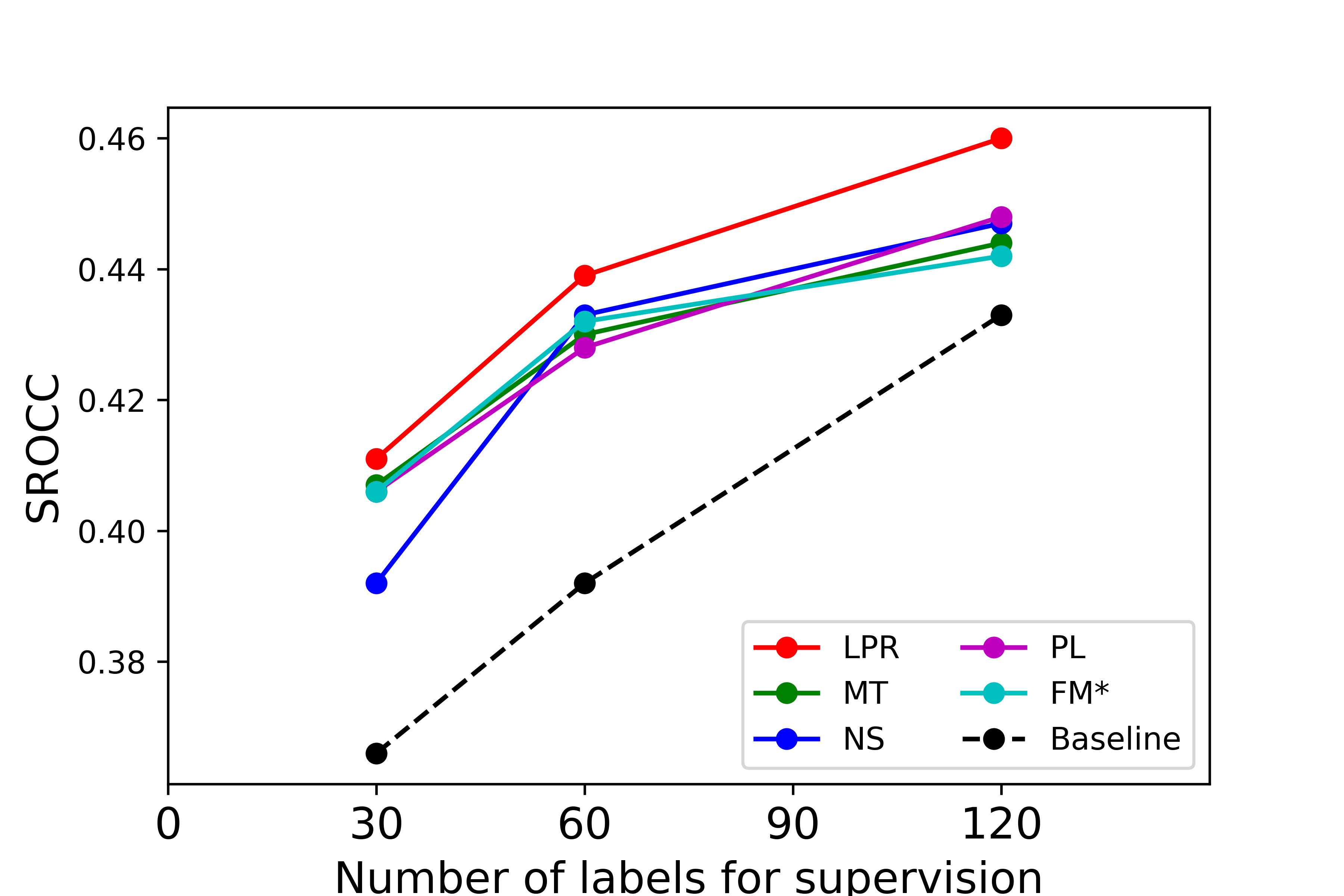}
         \caption{\footnotesize Train: KoNVid-1K; Test: LIVE Qualcomm}
     \end{subfigure}
     \hfill
     \begin{subfigure}[b]{0.32\textwidth}
         \centering
         \includegraphics[width=\textwidth]{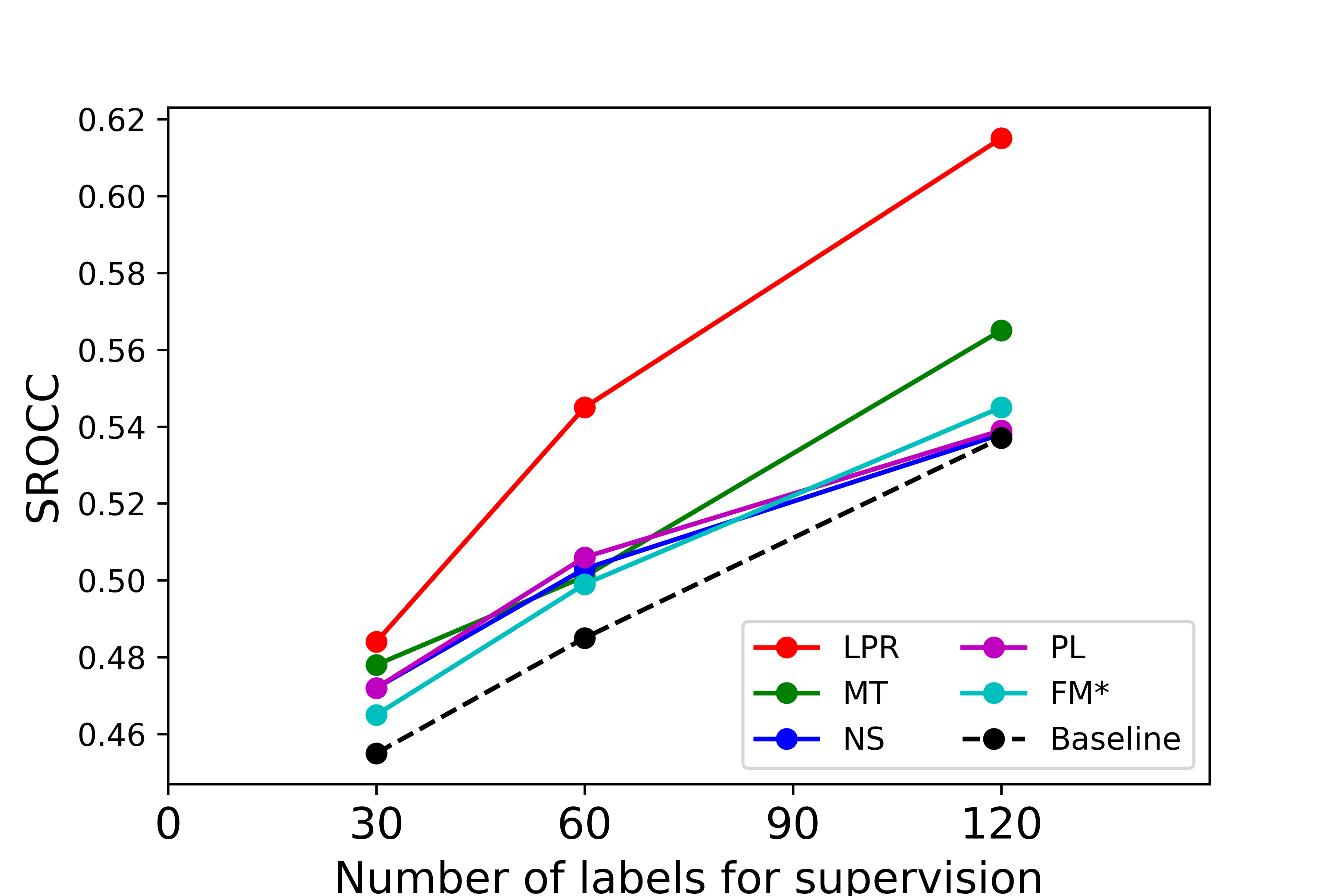}
         \caption{\footnotesize Train: LIVE VQC; Test: KoNVid-1K}
     \end{subfigure}\\

     \begin{subfigure}[b]{0.32\textwidth}
         \centering
         \includegraphics[width=\textwidth]{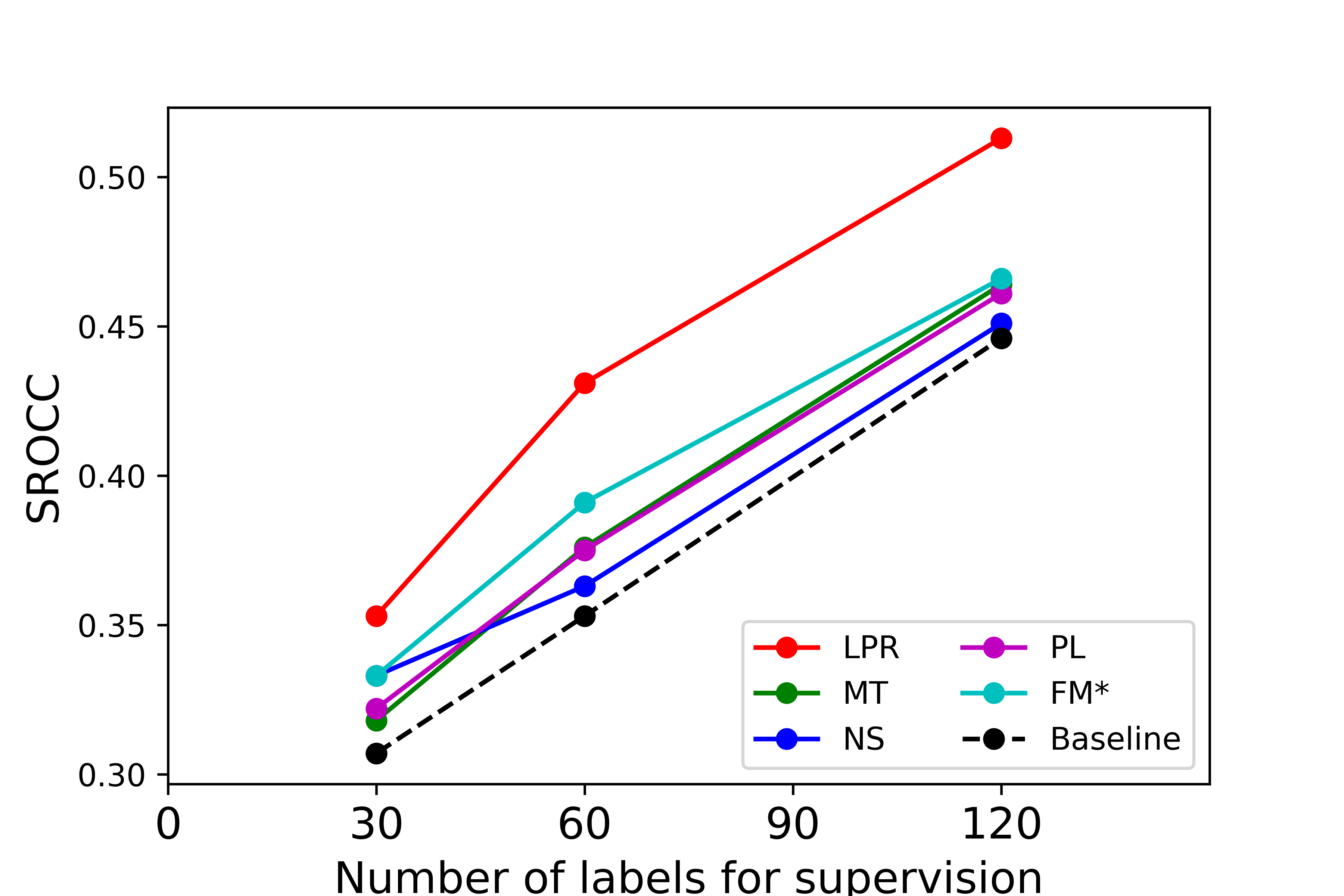}
         \caption{\footnotesize Train: LIVE VQC; Test: LIVE Qualcomm}
     \end{subfigure}     
     \hfill
     \begin{subfigure}[b]{0.32\textwidth}
         \centering
         \includegraphics[width=\textwidth]{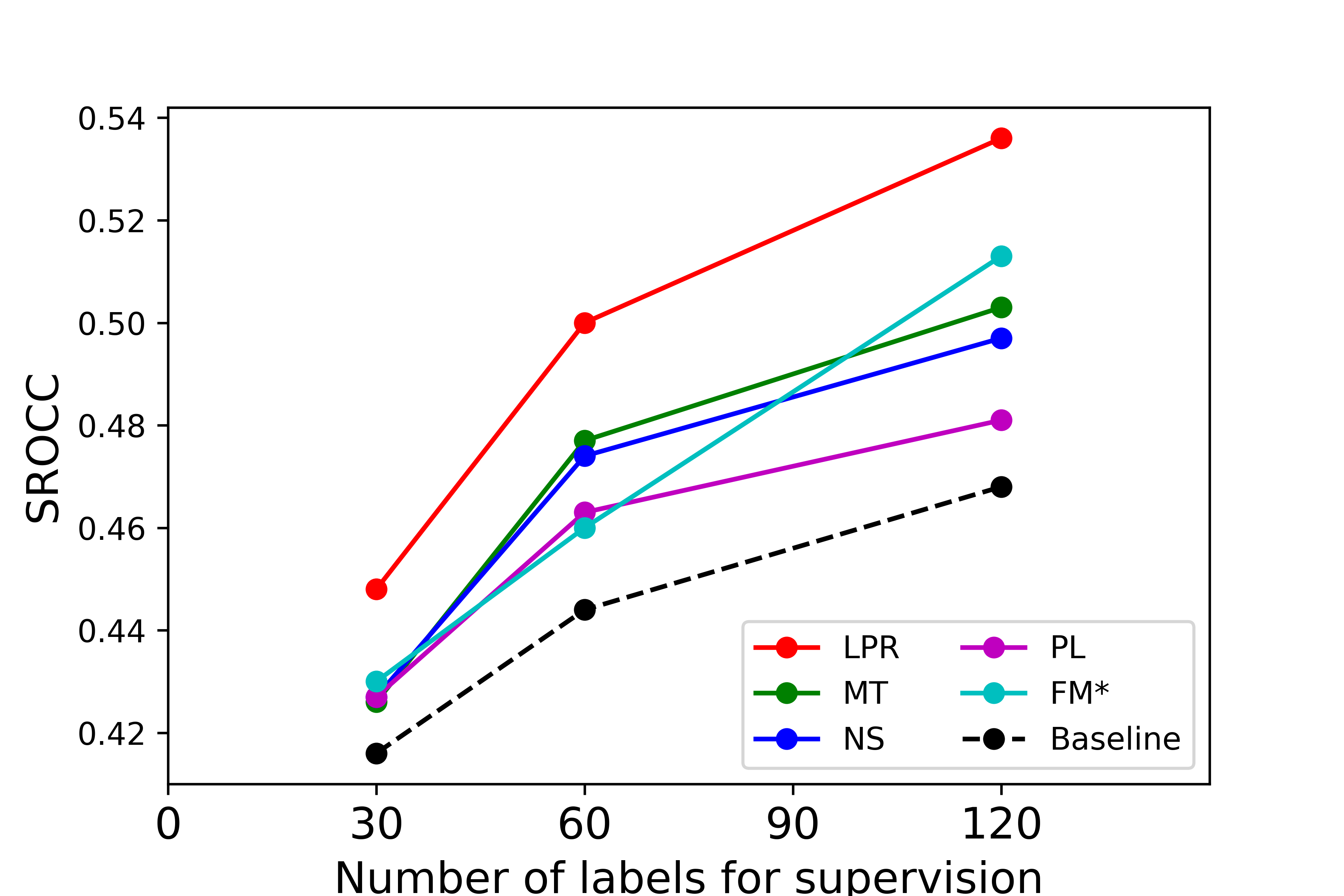}
         \caption{\footnotesize Train: LIVE Qualcomm; Test: KoNVid-1K}
     \end{subfigure}       
     \hfill
     \begin{subfigure}[b]{0.32\textwidth}
         \centering
         \includegraphics[width=\textwidth]{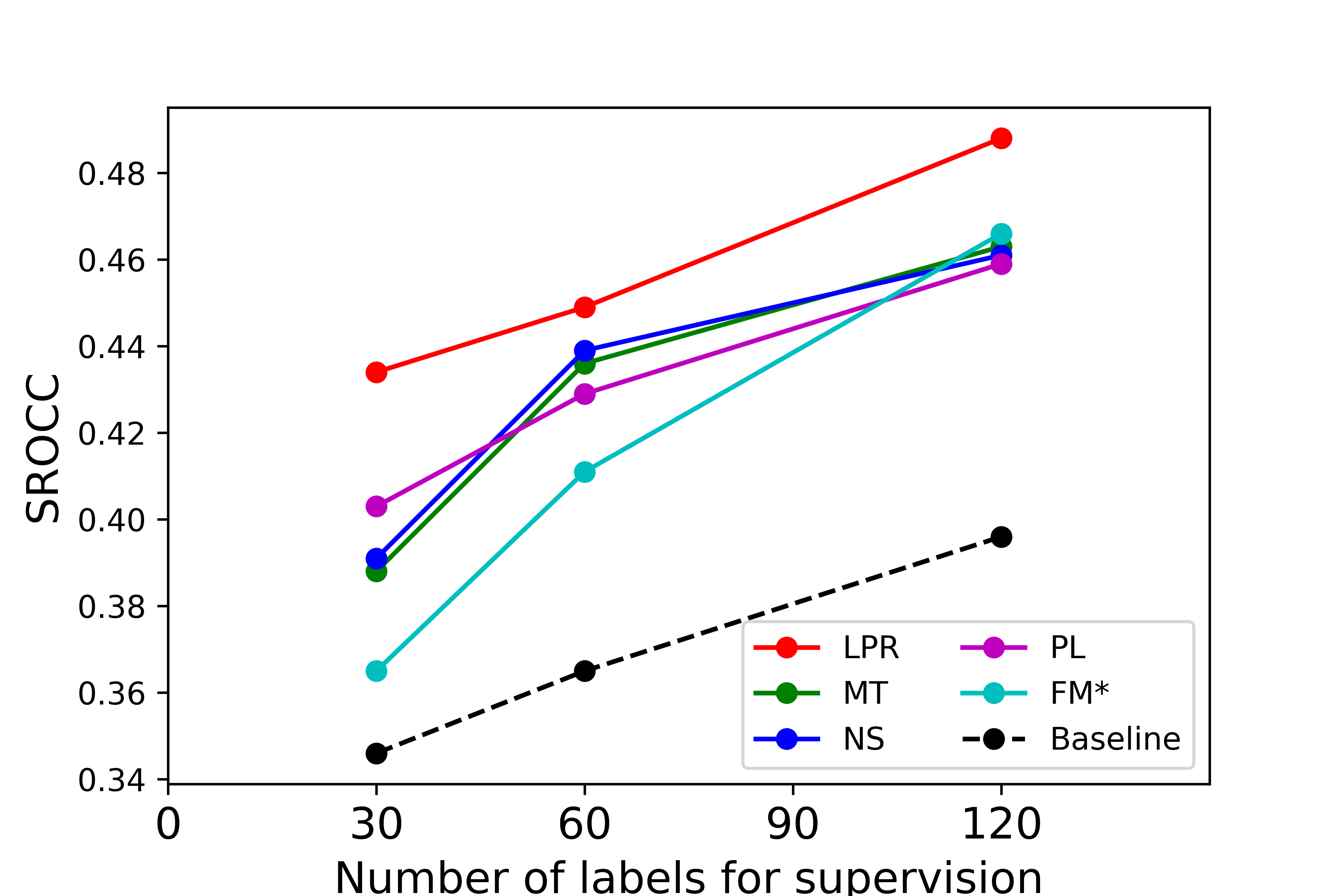}
         \caption{\footnotesize Train: LIVE Qualcomm; Test: LIVE VQC}
     \end{subfigure}  
     
        \caption{Cross database performance study of \textbf{LPR}, $FixMatch^*$ ($\textbf{FM}^*$), Mean Teacher (\textbf{MT}) \cite{mean_teach}, Noisy Student (\textbf{NS}) \cite{noisy_student}, and Pseudo-Labels (\textbf{PL}) \cite{pseudolabel}. The above six sub-figures show the comparative performance of the three aforementioned VQA algorithms under different train-test database settings. The line plots in each sub-figure correspond to the SROCC performance of each VQA model trained for $30$, $60$, and $120$ labels cases and tested on a different database.}
        \label{fig:crossdb_plot}
\end{figure*}

\begin{table*}
\caption{ SROCC performance analysis and comparison on KoNVid-1K, LIVE VQC, and LIVE Qualcomm datasets. The quality representation algorithms are regressed on 30, 60, and 120 labelled data respectively. The numbers in brackets indicate the increment in performance by learning on unlabelled data using LPR on various feature based algorithms.}
\centering
\label{tab:generalize}
\begin{adjustbox}{max width=\textwidth}
\begin{tabular}{|c|c|c|c|c|c|c|c|c|c|}
\hline
& \multicolumn{3}{c|}{KoNVid-1K} & \multicolumn{3}{c|}{LIVE VQC} & \multicolumn{3}{c|}{LIVE Qualcomm} \\
\hline
Algorithm & 30 labels & 60 labels & 120 labels & 30 labels & 60 labels & 120 labels & 30 labels & 60 labels & 120 labels \\
\hline 
TLVQM & 0.524 (\textcolor{red}{0.033}) & 0.599 (\textcolor{red}{0.023}) & 0.663 (\textcolor{red}{0.027}) & 0.588 (\textcolor{red}{0.047}) & 0.641 (\textcolor{red}{0.054}) & 0.663 (\textcolor{red}{0.053}) & 0.459 (\textcolor{red}{0.042}) & 0.577 (\textcolor{red}{0.033}) & 0.752 (\textcolor{red}{0.039})\\
VIDEVAL  & 0.513 (\textcolor{red}{0.050}) & 0.563 (\textcolor{red}{0.043}) & 0.617 (\textcolor{red}{0.024}) & 0.558 (\textcolor{red}{0.025}) & 0.634 (\textcolor{red}{0.061}) & 0.677 (\textcolor{red}{0.063}) & 0.488 (\textcolor{red}{0.086}) & 0.577 (\textcolor{red}{0.085}) & 0.663 (\textcolor{red}{0.092})\\
RAPIQUE & 0.549 (\textcolor{red}{0.051}) & 0.633 (\textcolor{red}{0.063}) & 0.694 (\textcolor{red}{0.059}) & 0.573 (\textcolor{red}{0.032}) & 0.659 (\textcolor{red}{0.048}) & 0.708 (\textcolor{red}{0.071}) & 0.452 (\textcolor{red}{0.081}) & 0.560 (\textcolor{red}{0.073}) & 0.661 (\textcolor{red}{0.077})\\
HEKE   & 0.516 (\textcolor{red}{0.053}) & 0.550 (\textcolor{red}{0.046}) & 0.623 (\textcolor{red}{0.057}) & 0.489 (\textcolor{red}{0.051}) & 0.540 (\textcolor{red}{0.050}) & 0.613 (\textcolor{red}{0.044}) & 0.442 (\textcolor{red}{0.067}) & 0.537 (\textcolor{red}{0.041}) & 0.623 (\textcolor{red}{0.040})\\
CNN-TLVQM  & 0.580 (\textcolor{red}{0.041}) & 0.670 (\textcolor{red}{0.038}) & 0.693 (\textcolor{red}{0.040}) & 0.591 (\textcolor{red}{0.058}) & 0.667 (\textcolor{red}{0.071}) & 0.686 (\textcolor{red}{0.068}) & 0.394 (\textcolor{red}{0.047}) & 0.578 (\textcolor{red}{0.035}) & 0.693(\textcolor{red}{0.038})\\
STED-TLVQM & 0.675 (\textcolor{red}{0.059}) & 0.708 (\textcolor{red}{0.043}) & 0.750 (\textcolor{red}{0.053}) & 0.621 (\textcolor{red}{0.060}) &0.709 (\textcolor{red}{0.051}) & 0.751 (\textcolor{red}{0.073}) & 0.557 (\textcolor{red}{0.082}) & 0.664 (\textcolor{red}{0.073}) & 0.794 (\textcolor{red}{0.029}) \\
\hline
\end{tabular}
\end{adjustbox}
\end{table*}

\subsubsection{Cross Database Performance Analysis}
 
 To analyze the generalization performance of our semi-supervised models, we conduct cross-database experiments and compare them with other SSL methods such as Mean Teacher \cite{mean_teach}, $FixMatch^*$, Noisy Student \cite{noisy_student}, and Pseudo-Labels \cite{pseudolabel}.  We take the models trained using a few labelled samples of one database and test them on a different database.  In total, we have six train-test settings across the KoNVid-1K \cite{konvid}, LIVE VQC \cite{livevqc}, and LIVE Qualcomm \cite{liveqcomm} databases. For each of these settings, we train the models for $30$, $60$, and $120$ label cases respectively, and report the results in Figure \ref{fig:crossdb_plot}. Our LPR model achieves superior performance compared to other semi-supervised algorithms in all the settings. We also provide the baseline performance in Figure \ref{fig:crossdb_plot} to show the relative gain in performance with learning with unlabelled videos for different SSL methods. We also note an improvement in SROCC values as the number of labels for supervision increases. This trend is in agreement with such improvement observed on the same database testing scenarios as well. 

\subsection{Learning Pseudo-Ranks with Different Quality Representations}

We now study the relevance of our LPR SSL model on various quality aware features described in Section \ref{QA_feature}. 
We note that both $f(\cdot)$, and $g(\cdot)$ are trained for frame level or hybrid model features, while for video level features $f(\cdot)$ does not exist. 
In Table \ref{tab:generalize}, we report the SROCC performance of LPR using the quality aware features under different settings of the number of labelled videos. We also show the increment in performance gain due to SSL by comparing it with the corresponding supervised learning with limited labels. We see that our SSL approach consistently improves the performance on all feature representations showing its stability.

\section{Ablation Studies}

\subsection{Impact of Various Components in LPR}
We first evaluate the need for each of the main contributions of our work in SSL, particularly, the need for augmentations, learning on pseudo-ranks and the threshold on the difference in the predicted scores of the teacher model to obtain reliable pseudo-ranks. 
In Table \ref{tab:ablation}, we report the results of an ablation experiment on KoNVid-1K, LIVE VQC, and LIVE Qualcomm databases for the 30 labelled videos scenario. 

\textbf{W/O Augmentation} We train LPR on the unlabelled samples without augmenting the input to either the student or teacher model. The student model is trained to predict the pseudo-ranks generated by the teacher. Note that the teacher is obtained as an exponential moving average of the student model parameters. 

\textbf{W/O Pseudo-Rank} In this setup, we modify the student-teacher model to learn pseudo-labels of the unlabelled videos rather than the pseudo-ranks of a video pair. Thresholding does not apply in this case as pseudo-labels of each unlabelled example are directly learnt here. This experiment proves the need for our generation of the pseudo-ranks. 

\textbf{W/O Thresholding} Here, the threshold $\tau$ is taken to be 0. Thus, the model is trained with all the unlabelled video pairs. This experiment studies the need for obtaining reliable pseudo-ranks by thresholding the difference in the quality predictions. 

\textbf{W/O Moving Average} In this experiment, we update the teacher weights with current student weights rather than the exponential moving average of past and present student weights. Here, the student and teacher models are identical and their outputs in the respective augmentations are required to be consistent. The gradients pertaining to the student model prediction loss are not propagated back to the teacher.

\begin{table}
\caption{Ablation Study}
\centering
\label{tab:ablation}
\begin{adjustbox}{max width=\columnwidth}
\begin{tabular}{|c|c|c|c|c|c|c|}
\hline
\multicolumn{4}{|c}{Model} & \multicolumn{3}{c|}{SROCC}\\ \hline
Moving  & Augment & Threshold & Rank & KoNVid-1K & LVQC & LQCOMM \\ 
Average & & & & & & \\ \hline
\checkmark & \checkmark & \checkmark & \checkmark   & 0.675 & 0.621 & 0.557\\
\hline
$\times$ & \checkmark & \checkmark & \checkmark & 0.648 & 0.595 & 0.497\\
\checkmark & $\times$ & \checkmark & \checkmark   & 0.642 & 0.587 & 0.504\\
\checkmark & \checkmark & $\times$ & \checkmark   & 0.631 & 0.578 & 0.492\\
\checkmark & \checkmark & $\times$ & $\times$   & 0.621 & 0.571 & 0.486\\
\hline
\end{tabular}
\end{adjustbox}
\end{table}

We observe from Table \ref{tab:ablation} that all the components of our model are important. However, we note that the generation of pseudo-ranks is extremely important and the performance drops significantly without this component. We observe that these trends are fairly consistent across all the datasets. 

\begin{figure*}
   \centering
   \begin{subfigure}[b]{0.32\textwidth}
     \centering
     \includegraphics[width=\textwidth]{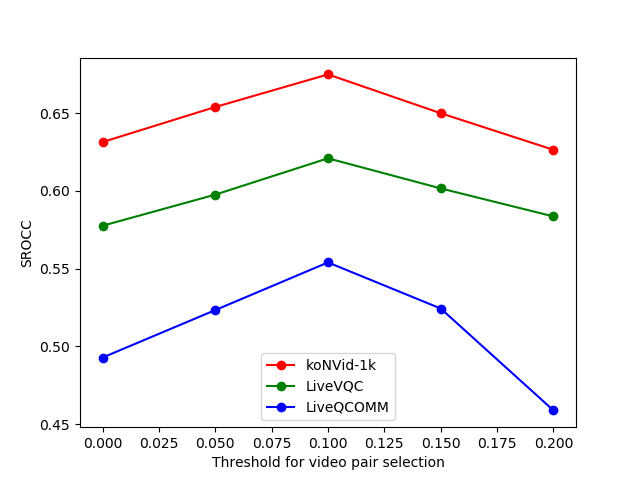}
     \caption{}
     \label{fig:threshold}
   \end{subfigure}
   \hfill
   \begin{subfigure}[b]{0.32\textwidth}
     \centering
     \includegraphics[width=\textwidth]{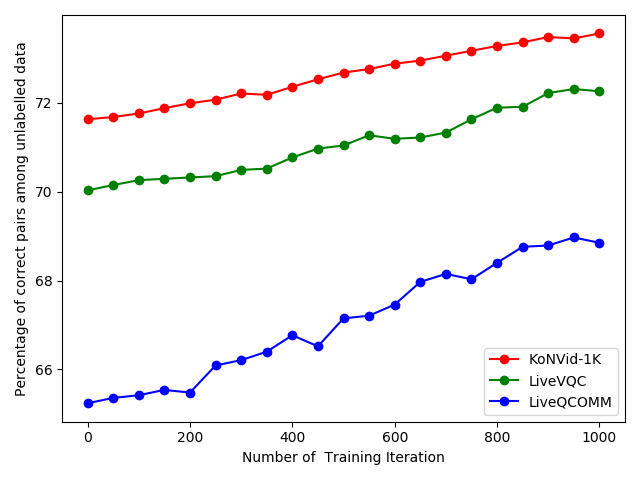}
     \caption{}
     \label{fig:unlabelled_accuracy}
   \end{subfigure}
   \hfill
\begin{subfigure}[b]{0.32\textwidth}
     \centering
     \includegraphics[width=\textwidth]{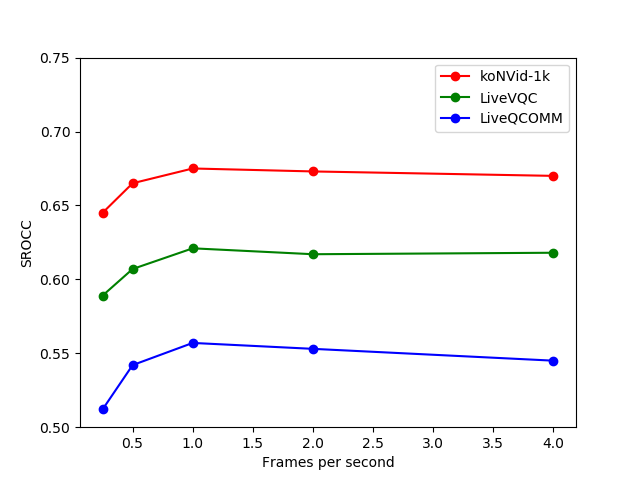}
     \caption{}
     \label{fig:fps}
   \end{subfigure}
   \hfill
    \caption{ (a) Performance analysis of LPR with threshold variation for the 30 MOS labels case (b) Percentage of video pairs with correct rank order with training iterations for LPR in the 30 MOS labels case (c) Performance variation of LPR with different frame rate was chosen for strongly augmenting the features.}
\end{figure*}

\subsection{Analysis of Hyper Parameters} 

We now analyze our model with respect to the threshold $\tau$, and how the performance on the unlabelled data improves with training. In Figure \ref{fig:threshold}, we first analyze how the performance (SROCC) of our model varies with respect to the choice of the threshold $\tau$. We present the results when 30 labels are available in the respective datasets. While a threshold $\tau=0$ implies that all the pseudo-rank pairs are selected, a very high choice of the threshold implies very few pairs are selected. We observe that a choice of $\tau=0.1$ yields an optimum performance across different datasets. While the drop in performance is steady for $\tau>0.1$ on the LIVE VQC, and KoNViD-1K datasets, there is a steeper drop in performance on the LIVE Qualcomm database. The LIVE Qualcomm database is a smaller database, and with larger values of $\tau$, very few unlabelled videos that satisfy the threshold criterion are selected, which can bias the training, and lead to poorer performance. 

In Figure \ref{fig:unlabelled_accuracy}, we track the performance of our model on the unlabelled data as training proceeds. This experiment aims to understand how the model improves with training, ultimately leading to its superior performance on the testing dataset. We evaluate the accuracy of the pseudo-ranks for unlabelled video pairs with respect to the true ranks for these pairs. In all three datasets, we see that as learning proceeds, the fraction of unlabelled video pairs with the correct pseudo-ranks keep improving. Nevertheless, some saturation is seen as the training proceeds beyond a point. 

The student network is fed with strongly augmented features with severe subsampling of the STED features as input. In Figure \ref{fig:fps}, we vary these frame rates from $0.25$ to $4$ frames per second and record the performance. We see that sub-sampling the input of less than 1 frame per second (fps) impacts the performance due to a significant reduction in the frame level feature information. As the performance is fairly consistent between 1-4 fps across all the datasets, we choose a sub-sampling rate of 1 fps to strongly input the frame level quality features.  Note that as the frame rate increases, the effect of the strong-weak augmentation decreases. 

\begin{figure*}
     \centering
     \begin{subfigure}[b]{0.32\textwidth}
         \centering
         \includegraphics[width=\textwidth]{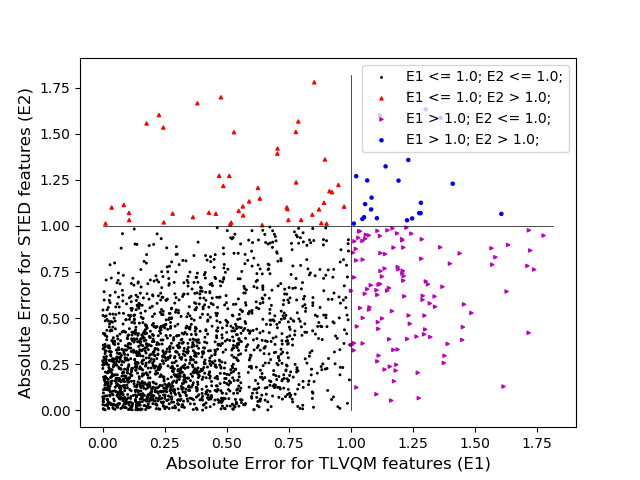}
         \caption{\footnotesize KoNVid-1K}
     \end{subfigure}
     \hfill
     \begin{subfigure}[b]{0.32\textwidth}
         \centering
         \includegraphics[width=\textwidth]{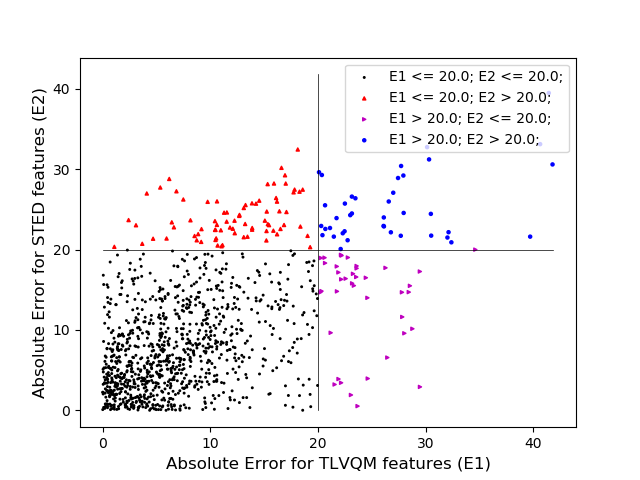}
         \caption{\footnotesize LIVE VQC}
     \end{subfigure}
     \hfill
     \begin{subfigure}[b]{0.32\textwidth}
         \centering
         \includegraphics[width=\textwidth]{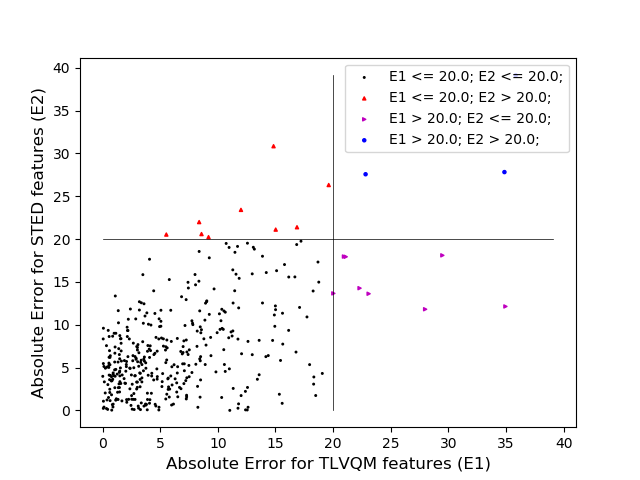}
         \caption{\footnotesize LIVE Qualcomm}
     \end{subfigure}
        \caption{Analysis of complementarity between STED features and TLVQM \cite{tlvqm} features}
        \label{fig:complementary}
\end{figure*}

\subsection{Complementarity of quality aware features} We perform an error-based complementarity study on the STED, and TLVQM features in predicting the video quality scores. We regress the TLVQM \cite{tlvqm} features, and STED based features against the ground truth MOS on $80\%$ data for each of the three authentically distorted video databases, respectively. We then compute the absolute error between the predicted quality, and the MOS of the remaining $20\%$ test videos. Figure \ref{fig:complementary} shows the scatter plot between the absolute error in predicting the quality using the TLVQM \cite{tlvqm}, and STED features. The plot is divided into four quadrants based on whether the individual error is greater or less than a $20\%$ of the MOS range of that particular dataset. We see that the two sets of features give complementary predictions for certain samples, and thus combining them can improve the overall model performance as evident from Tables \ref{tab:feat_srocc} and \ref{tab:feat_lcc}.


\section{Conclusion}
We designed a framework for NR VQA of authentically distorted videos when only limited labels are available for training a video quality model. We showed the effective use of the unlabelled videos by generating pairwise pseudo-ranks with student-teacher models on strong-weak augmented videos, and using such ranks to improve the model. While we showed the utility of our learning approach on different features, we also presented a particular feature model for spatial, and temporal features learned with spatio-temporal entropic differences. 
Our framework shows that one can significantly improve the performance on authentically distorted videos in terms of correlation with human perception, even when only a few videos are labelled with human opinion scores. 

\bibliographystyle{ACM-Reference-Format}
\bibliography{paper}

\end{document}